\documentclass[aps,twocolumn,showpacs,amsmath,amssymb,amsfonts]{revtex4}

\usepackage{graphicx}

\newcommand{\ket}[1]{\, | #1 \rangle}
\newcommand{\braket}[2]{\langle #1 | #2 \rangle}
\newcommand{\la}{\lambda}

\newcommand{\De}{\Delta}

\newcommand{\veps}{\varepsilon}
\newcommand{\us}{\uparrow}
\newcommand{\ds}{\downarrow}
\newcommand{\lra}{\leftrightarrow}
\newcommand{\adg}{a^{\dagger}}

\begin{document}

\title{Coherent population transfer in a chain of tunnel coupled
quantum dots\footnote{This paper is dedicated to Bruce W. Shore 
on the occasion of his 70th birthday.}}

\author{D. Petrosyan}
\affiliation{Institute of Electronic Structure \& Laser, 
FORTH, 71110 Heraklion, Crete, Greece}
\author{P. Lambropoulos}
\affiliation{Institute of Electronic Structure \& Laser, 
FORTH, 71110 Heraklion, Crete, Greece}

\begin{abstract}
We consider the dynamics of a single electron in a chain of tunnel 
coupled quantum dots, exploring the formal analogies of this system with 
some of the laser-driven multilevel atomic or molecular systems 
studied by Bruce W. Shore and collaborators over the last 30 years.
In particular, we describe two regimes for achieving complete coherent
transfer of population in such a multistate system. In the first regime,
by carefully arranging the coupling strengths, the flow of population
between the states of the system can be made periodic in time. In the 
second regime, by employing a ``counterintuitive'' sequence of couplings,
the coherent population trapping eigenstate of the system can be rotated
from the initial to the final desired state, which is an equivalent of 
the STIRAP technique for atoms or molecules. Our results may be useful
in future quantum computation schemes.
\end{abstract}


\maketitle

\section{Introduction}

Population transfer in multistate quantum systems has been an active 
topic of research over the last half a century. In the context of atomic
and molecular physics, coherent population transfer in optically-driven 
multilevel systems has been studied since the invention of lasers 
\cite{shoreBook}. Usually, the objective is to transfer the population
from the initial to a well defined final state of the atom or molecule,
via one or more intermediate states, while minimizing the loss of 
population through or its accumulation on the intermediate states. 
In early theoretical work, Shore and collaborators have studied 
population transfer in multilevel systems driven by resonant laser 
fields \cite{ShrEbr,CookShore,Shore}. In particular, they have found 
that it is possible to arrange the coupling strengths between the 
adjacent states in such a way that the system becomes analogous 
to a spin-$J$ in a magnetic field, whose dynamic evolution is known 
to be periodic for any $J$ \cite{CookShore}. This coupling scheme 
was therefore named spin-coupling.

Later, Hioe, Eberly, Bergmann and collaborators discovered the technique 
of stimulated Raman adiabatic passage (STIRAP) for three-level 
atomic/molecular systems \cite{stirap3ls}. They have identified a specific 
eigenstate of the system, the so-called coherent population trapping (CPT) 
state, which contains a superposition of the initial and final states, 
and dates back to Alzetta et al. and Arimondo and Orriols \cite{CPTeth}. 
The STIRAP technique is then based on first preparing the system in 
its initial bare state, which coincides with the CPT state, and then 
adiabatically rotating the CPT state towards the desired final bare 
state of the system. This techniques has been subsequently polished 
\cite{stirap-rev} and extended to multilevel systems 
\cite{stirap4ls,stirapN-DT,stirapNsqLs} with the active participation 
of Bruce W. Shore. 

While the above studies were conducted in the context of multilevel atoms 
or molecules, here we show that similar effects can be found in the 
context of quantum transport in arrays of tunnel-coupled quantum dots 
\cite{QDarray,DasSarma,WeNa,GCHH}. Often referred to as artificial atoms, 
semiconductor quantum dots offer an unprecedented possibility of constructing
at will and exploring situations ranging from practically single atom to a 
fully solid state many-body systems \cite{QDrev}. The nanofabrication 
possibilities of tailoring structures to desired geometries and 
specifications, and controlling the number and mobility of electrons 
confined within a region of space, are some of the features that make 
these structures unique tools for the study {of a} variety of preselected
set of phenomena, including the coherent population transfer in multistate
systems.

Given the controllable quantum properties of the electrons in such structures,
the possibility of their application to schemes of quantum computers (QCs) 
\cite{QCI} has not escaped attention \cite{LDV,QDQCdsgn,zanros}. The 
qubits of the QD-array based QC would be represented by the spin-states 
of single electrons confined in individual QDs, with the two-qubit 
nearest-neighbor coupling mediated by the controlled spin-exchange 
interaction \cite{LDV,QDQCdsgn}. One of the main difficulties with the 
existing proposals for integrated solid-state based QCs is that there is no 
efficient way of transferring the information between distant qubits.
We consider here a single-electron tunneling in a one-dimensional array
of QDs and establish the conditions under which the complete transfer of 
the electron wavepacket between two distant locations can be achieved. 
Our findings could therefore be relevant to the reliable information exchange 
between distant parts of an integrated quantum computer \cite{weNPL}. 

In Section \ref{sec:mform} we outline the mathematical formalism 
describing a chain of QDs, in terms of which, in Section \ref{sec:spin}, 
we present the theory of coherent propagation and periodic oscillations 
of the electron wavepacket between the two ends of the chain. The 
single-electron transfer via an equivalent of multistate STIRAP is 
discussed in Section \ref{sec:stirap}. In Section \ref{sec:concl} we 
describe an envisioned implementation of a scalable quantum computer, 
followed by the concluding remarks.

\section{Mathematical formalism}
\label{sec:mform}

We consider electron transport in a linear array of $N$ nearly 
identical QDs which are electrostatically defined in a two-dimensional 
electron gas by means of metallic gates on top of a semiconductor 
heterostructure (GaAs/AlGaAs) \cite{QDarray,QDrev}. This system is 
described by the extended Mott-Hubbard Hamiltonian \cite{DasSarma,WeNa}, 
which in its most general form is given by
\begin{eqnarray}
H &=& \sum_{j,\alpha} \veps_{j\alpha} \adg_{j\alpha} a_{j\alpha}
+\frac{1}{2} \sum_{j} U n_j (n_j -1) \nonumber \\ & &
+\sum_{i < j,\alpha} t_{ij,\alpha} (\adg_{i\alpha} a_{j\alpha}+ 
a_{i\alpha} \adg_{j\alpha}) + \sum_{i < j} V_{ij} n_i n_j , \label{Ham}
\end{eqnarray}
where $\adg_{j\alpha}$ and $a_{j\alpha}$ are the creation and 
annihilation operators for an electron in state $\alpha$ with the 
single-particle energy $\veps_{j\alpha}$, $U$ is the on-site Coulomb repulsion,
$n_j=\sum_{\alpha} \adg_{j\alpha} a_{j\alpha}$ the total electron number 
operator of the $j$th dot, $t_{ij,\alpha}$ are the coherent tunnel 
matrix elements between dots $i$ and $j$, and $V_{ij}$ is the interdot 
electrostatic interaction. In general, the index $\alpha$ refers to 
both orbital and spin states of an electron. In the tight-binding regime, 
when the on-site Coulomb repulsion and single-particle level-spacing 
$\De \veps$ are much larger than the tunneling rates, 
$U > \De \veps \gg t_{ij,\alpha}$, only the equivalent states 
of the neighboring dots are tunnel-coupled to each other \cite{cmnt}.
In the absence of a magnetic field, we can thus limit our consideration 
only to a single doubly- (spin-) degenerate level per dot 
($\alpha\in\{\us,\ds\}$), assuming further that the tunneling rates 
do not depend on the electron spin.

\begin{figure}[t]
\centerline{\includegraphics[width=6cm]{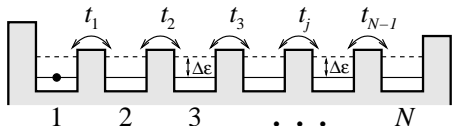}}
\caption{Schematic drawing of the chain of tunnel-coupled QDs.
\label{qdch_1e}}
\end{figure}

In this paper we are concerned with single-electron dynamics, 
considering a situation in which a preselected QD is initially 
doped with one mobile electron, while all of the other dots of the chain
are empty, as indicated in Fig.~\ref{qdch_1e}. Our aim is to determine 
the conditions under which the complete coherent transfer of the 
electron between the two ends of the chain can be achieved.
The population transfer in this system is mediated by the tunneling
between the neighboring QDs. The individual tunneling rates 
$t_j \equiv t_{jj+1}$ are determined by the voltages applied to 
the gates defining the corresponding interdot tunneling barriers. 
A chain of $N$ tunnel-coupled QDs doped with a single electron is 
described by the following Hamiltonian,
\begin{equation}
H_{\rm 1e} = \sum_{j,\alpha} \veps_{j} \adg_{j\alpha} a_{j\alpha}
+ \sum_{j,\alpha} t_j (\adg_{j \alpha} a_{j+1,\alpha}+ 
a_{j\alpha} \adg_{j+1,\alpha}) \label{Ham1e},
\end{equation}
which obviously does not contain terms responsible for electrostatic
interactions. Since this Hamiltonian preserves the electron number and 
its spin, the total state-vector of the system reads 
\begin{equation}
\ket{\psi(\tau)} = \sum_{j,\alpha}^N A_{j}^{\alpha}(\tau)\ket{j_\alpha}
\label{wfunct},
\end{equation} 
where $\ket{j_\alpha} \equiv \adg_{j\alpha} \ket{0_1,...,0_N}$ denotes 
the state with one electron having spin $\alpha$ at the $j$th dot. The 
time-evolution of the system is governed by the Schr\"odinger equation 
$\mathrm{i}\ket{\dot{\psi}} = H_{\rm 1e}\ket{\psi}$ ($\hbar =1$), which 
yields 
\begin{equation}
\mathrm{i} \frac{d A_j^{\alpha}}{d \tau} = \veps_{j} A_j^{\alpha} 
+ t_{j-1} A_{j-1}^{\alpha} + t_{j} A_{j+1}^{\alpha} , 
\label{em1e}
\end{equation} 
where $t_{0} = t_{N} =0$. Obviously, the two sets of these amplitude 
equations with $\alpha = \us$ and $\alpha = \ds$ are equivalent and 
decoupled from each other. As a result, if the electron is prepared 
in an arbitrary superposition of spin up and spin down states, 
$\ket{\psi} = A_{j}^{\us}\ket{j_\us}+A_{j}^{\ds}\ket{j_\ds}$, the two 
parts of the wavefunction evolve symmetrically and independently of 
each other. This assertion is valid as long as all the uncontrollable
spin-flip processes are vanishingly small on the time scale of $t^{-1}$.
In semiconductor QDs, the spin decoherence originates mainly from the 
spin-phonon coupling, as well as the coupling of the electron spin 
with the nuclear spins of the surrounding crystal (hyperfine interaction)
or stray magnetic fields. The first decoherence mechanism is suppressed 
at low temperatures \cite{cmnt}, at which the density of crystal phonons 
is negligible \cite{phonons}. As for the uncontrollable hyperfine 
interactions, experimental measurements indicate spin-relaxation times
in excess of $100\:\mu$s, which can be further improved by applying
moderate magnetic fields or polarizing the nuclear spins \cite{sRLX}.
Another mechanism for decoherence in the process of electron (charge)
transfer in our system originates from the structure imperfections and
gate voltage fluctuations, which cause uncertainty in the intradot energy 
levels and interdot couplings. These fluctuations, however, are typically
slow on the time scale of $t^{-1}$, and the resulting disorder in the
system may be considered frozen during its dynamic evolution, as we 
have discussed in a previous publication \cite{weNPL}.  

Let us write the Hamiltonian for the electron with spin $\alpha$
in the matrix form
\begin{equation}
H_{\rm 1e}^{\alpha} = 
\left[ \begin{array}{cccccc}
\veps_1 &   t_1   &    0    & \cdots &     &    \\ 
  t_1   & \veps_2 &    t_2  &        &     &     \\
   0    &   t_2   & \veps_3 &        &     &    \\
\vdots  &         &         & \ddots &     &  \vdots \\
        &         &         &        & \veps_{N-1}  & t_{N-1}  \\ 
        &         &         & \cdots &  t_{N-1} & \veps_N 
\end{array} \right] , \label{Ham1ealpha}
\end{equation} 
which is obviously tridiagonal. Inspection of the amplitude 
equations (\ref{em1e}) or the Hamiltonian (\ref{Ham1ealpha}) indeed 
verifies that our system is formally analogous to the laser-driven 
multilevel atomic or molecular systems studied by Shore and coworkers 
\cite{ShrEbr,CookShore,Shore} and Bergmann, Shore and others 
\cite{stirap3ls,stirap-rev,stirap4ls,stirapN-DT,stirapNsqLs}. 
Here, the tunneling rates $t_j$ between states $\ket{j}$ and $\ket{j+1}$
play the same role as the Rabi frequencies of the laser fields acting
on the atomic transitions $\ket{j} \lra \ket{j+1}$, while the energies
$\veps_{j}$ of states $\ket{j}$ correspond to the cumulative detunings of 
the atomic levels. In the following Sections, we describe two methods for
achieving complete population transfer from the initial $\ket{1}$ to the 
final $\ket{N}$ state of the system, which turn out to be the counterpart
of those in Refs. \cite{CookShore} and \cite{stirapNsqLs}.

\section{Periodic oscillations of population between the two end states}
\label{sec:spin}
 
In this Section we consider the electron wavepacket dynamics in the 
chain with static couplings between the dots. Assume that at time 
$\tau=0$ the electron is localized on the first dot, 
$\ket{\psi^{\alpha}(0)} = \ket{1_{\alpha}}$, and the tunnel couplings 
are switched on. This switching should be fast enough on the time scale 
of $t^{-1}$, so that no appreciable change in the initial state of the
system occurs during the switching time $\tau_{\rm sw}$, but slow on 
the time scale of $\veps^{-1}$, so that no nonresonant coupling
between the dots is induced: $\veps^{-1} < \tau_{\rm sw} < t^{-1}$. The 
aim is to determine the set of couplings between the states of the systems 
which will achieve a complete transfer of the electron population from 
the initial to the final dot. 

\begin{figure*}[t]
\centerline{\includegraphics[width=11cm]{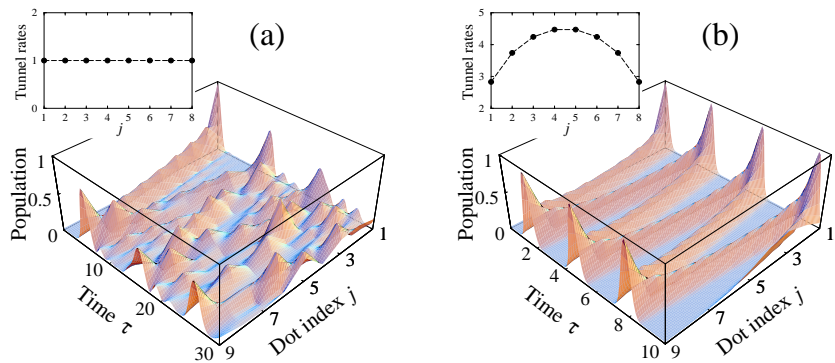}}
\caption{Time-evolution of a single-electron wavepacket in a chain of 
$N=9$ QDs with static tunneling rates. 
(a) Population flow in the chain with equal interdot tunneling rates
$t_j = t$ (shown in the inset). 
(b) Population flow in the chain with spin-model tunneling rates 
$t_j = t \sqrt{(N-j)j}$ (shown in the inset).
The time $\tau$ is in units of $t^{-1}$.
\label{qdc_1e_St}}
\end{figure*}

To determine the time-evolution of the state vector (\ref{wfunct})
we need to solve the eigenvalue problem 
$H_{\rm 1e}^{\alpha} \ket{\psi^{\alpha}} = \la \ket{\psi^{\alpha}}$
which will yield the eigenvalues $\la_k$ and corresponding 
eigenvectors $\ket{\psi_k^{\alpha}}$ of the Hamiltonian (\ref{Ham1ealpha}).
The state vector $\ket{\psi^{\alpha}(\tau)}$ at any time $\tau \geq 0$
is given by
\begin{equation}
\ket{\psi^{\alpha}(\tau)} = \sum_k^N e^{- \mathrm{i} \la_k \tau}
\ket{\psi_k^{\alpha}} \braket{\psi_k^{\alpha}}{\psi^{\alpha}(0)}
= \sum_{j}^N A_{j}^{\alpha}(\tau) \ket{j_\alpha} . \label{EigstExp}
\end{equation}
Note that the matrix in Eq. (\ref{Ham1ealpha}) has the form of the 
tridiagonal Jacobi matrix. It is natural to first consider the case 
of equal tunneling rates between the dots: $t_j = t$. Assuming 
equal energies $\veps_j = \veps$ and making the transformation 
$A_j^{\alpha}\to A_j^{\alpha} e^{i \veps \tau}$, which is equivalent
to the interaction picture, we find that the determinant 
$\mathcal{D}_N(\la) \equiv \det(H_{\rm 1e}^{\alpha} - \la \mathbb{I})$ 
is identical to the Chebyshev polynomial of the second kind, which can
be expressed as $\mathcal{D}_N(\la) = \Pi_{k=1}^{N} (\la-\la_k)$. 
The eigenenergies of the system are then given by the roots of this 
polynomial, namely
\[
\la_k = 2 t \cos \left( \frac{k \pi}{N+1}\right) ,
\] 
while the corresponding eigenvectors are
\[
\ket{\psi_k^{\alpha}} = \sqrt{\frac{2}{N+1}} \sum_{j}^N
\sin\left( \frac{j k \pi}{N+1}\right) \ket{j_{\alpha}} .
\]
Using Eq. (\ref{EigstExp}) and the initial conditions $A_1 = 1$ and 
$A_j = 0$ for $j=2,3,\ldots N$, we obtain the solutions for the 
amplitudes as, 
\begin{eqnarray}
A_j^{\alpha} &=&  \frac{2}{N+1} \sum_{k=1}^{N} 
\exp \left[-\mathrm{i} 2 t \tau \cos \left( \frac{k \pi}{N+1}\right)\right] 
\nonumber \\
& & \times \sin\left( \frac{j k \pi}{N+1}\right) 
\sin\left( \frac{k \pi}{N+1}\right).
\end{eqnarray}
It is thus evident that the eigenstates of the coupled system oscillate 
with incommensurate frequencies corresponding to the roots $\la_k$ of 
$\mathcal{D}_N$, which in fact become increasingly densely spaced with 
increasing $N$. As a consequence, the system never revives fully to its 
initial state, as is illustrated in Fig.~\ref{qdc_1e_St}(a).

Clearly, it is highly desirable to tailor the parameters of the system
so as to achieve a non-dispersive transfer of the single-electron 
wavepacket between the two ends of the chain. Recall from the theory 
of angular momentum that a spin-$J$ particle subject to a constant 
magnetic field exhibits Larmor precession about the field direction.
In particular, if one chooses the quantization direction along an axis
perpendicular to the magnetic field direction and prepares the particle
in its lowest spin eigenstate $\ket{J,M=-J}$, it will oscillate between
this initial and the final state $\ket{J,M=J}$ in a perfectly periodic 
way. The matrix elements for the transitions $\ket{J,M} \lra \ket{J,M+1}$
between the neighboring states are proportional to $\sqrt{(J-M)(J+M+1)}$.
It is therefore clear that with the appropriate choice of the interdot 
tunneling matrix elements, the dynamics of the single-electron in a chain
of QDs can mimic that of a spin-$J$ in a magnetic field. Indeed, if we 
formally set $N = 2J+1$ and $j = J+ M +1$, the tunneling rates $t_j$ 
should be arranged according to $t_j=t \sqrt{(N-j)j}$ for $j=1,...,N-1$. 
Then again, by exploring the properties of the Jacobi polynomials, 
we find equally spaced eigenenergies of the system,
\[
\la_k = t (2k-N-1) ,
\]
while the corresponding eigenvectors can be expressed through the rotation
matrices commonly used in the representation theory of angular momentum. 
With the initial conditions $A_1 = 1$ and $A_j = 0$ for $j=2,3,\ldots N$, 
for the amplitudes of the state-vector (\ref{wfunct}), we then obtain 
simple analytic expressions given by the binomial form
\begin{equation}
A_j^{\alpha} = \left(\begin{array}{c}
N-1 \\ j-1 \end{array} \right)^{1/2} 
[-\mathrm{i} \sin{(t \tau)}]^{(j-1)} \cos{(t \tau)}^{(N-j)} .
\end{equation}
Since the eigenstates of the system have commensurate energies $\la_k$,
the electron wavepacket oscillates in a perfectly periodic way between
the first and the last dots, whose occupation probabilities are given, 
respectively, by $|A_1^{\alpha}|^2 = \cos{(t \tau)}^{2(N-1)}$ and 
$|A_N^{\alpha}|^2 = \sin{(t \tau)}^{2(N-1)}$, which is illustrated 
in Fig.~\ref{qdc_1e_St}(b). In particular, if at time 
$\tau = \pi /(2 t)$ the tunneling rates are suddenly switched off,
we obtain $|A_1^{\alpha}|^2 = 0$ and $|A_N^{\alpha}|^2 = 1$, i.e. 
complete population transfer from the initial to the final state 
of the system. In a somewhat abstract sense, the behavior of the 
system is thus similar to that of a two-level system subject to a 
$\pi$ pulse.  Let us note at this point that the population transfer
between the two ends of the chain can be achieved most straightforwardly
by sequentially pulsing the tunneling rates between the first and 
second dots for time $\tau_1 = \pi/(2 t_1)$, then the second and third
dots for time $\tau_2 = \pi/(2 t_2)$, etc till reaching the $N$th dot, 
which is equivalent to applying a sequence of $\pi$ pulses in a multistate
atomic system. In the scheme described above, however, all the interdot
tunnelings are switched on and then off simultaneously, realizing 
thereby a fast and efficient transfer of the electron from the first 
to the last QD.

\section{Adiabatic population transfer between the two end states}
\label{sec:stirap}

While the above tunneling schemes, involving a sequence of $\pi$ pulses or 
an effective collective $\pi$ pulse, require both, careful control of the 
individual tunneling rates and their timing, in this Section we describe
a robust adiabatic method for population transfer which is not very 
sensitive to small uncertainties in the interdot tunneling rates.
Recall that a three-level atom interacting with two laser fields, 
under the condition of two-photon (Raman) resonance, possesses a coherent
population trapping (CPT) state, which is decoupled from both laser fields 
\cite{stirap-rev}. Equivalently, for a chain of three tunnel-coupled 
quantum dots, assuming equal energies $\veps_j = \veps$, the eigenstate 
of Hamiltonian (\ref{Ham1ealpha}) with zero eigenvalue, $\la_0=0$, is 
given by 
\begin{equation}
\ket{\psi_0^{\alpha}} = \frac{1}{\sqrt{\mathcal{N}_0}} 
[t_2 \ket{1_{\alpha}} - t_1 \ket{3_{\alpha}}],  \qquad
\mathcal{N}_0 = t_1^2 + t_2^2 \label{CPT3l}.
\end{equation}
This is a CPT state that does not contain a contribution from the intermediate
state $\ket{2_{\alpha}}$. The other two eigenstates
\begin{eqnarray*}
\ket{\psi_{\pm}^{\alpha}} &=&  \frac{1}{\sqrt{\mathcal{N}_{\pm}}}
[t_1\ket{1_{\alpha}} - \la_{\pm} \ket{2_{\alpha}} + t_2 \ket{3_{\alpha}}], \\
& & \mathcal{N}_{\pm} = t_1^2 + \la_{\pm}^2 + t_2^2 = 2 \mathcal{N}_0 ,
\end{eqnarray*}
with corresponding eigenvalues $\la_{\pm} = \pm \sqrt{t_1^2 + t_2^2}$,
contain all three states $\ket{j_{\alpha}}$. If for a given coupling 
strengths $t_1$ and $t_2$ the system is prepared in the CPT state 
(\ref{CPT3l}), it will remain in this state as long as the couplings are 
constant in time. But even for time-dependent couplings, the system 
initially prepared in the CPT state can adiabatically follow this state, 
provided the tunneling rates change slowly enough. More quantitatively,
the nonadiabatic coupling between the eigenstates of Hamiltonian 
(\ref{Ham1ealpha}) is small, if during the evolution the transition 
amplitude $\braket{\psi_{\pm}^{\alpha}}{\dot{\psi}_0^{\alpha}}$ remains
much smaller than the energy separation between the corresponding eigenstates 
\cite{stirap-rev},
\begin{equation}
|\braket{\psi_{\pm}^{\alpha}}{\dot{\psi}_0^{\alpha}}| \ll |\la_{\pm} - \la_0|.
\label{adiabat}
\end{equation}
Our objective is to transfer the electron from the first to the last QD 
using the time-dependent (pulsed) tunnel-couplings. From Eq. (\ref{CPT3l})
one can see that if at an early time the tunnel coupling $t_2$ is switched
on while $t_1 \ll t_2$, the CPT state coincides with the initial state
$\ket{1_{\alpha}}$. One then slowly (adiabatically) decreases $t_2$ while
increasing $t_1$, so that at a later time  $t_1 \gg t_2$ and the CPT state 
coincides with the final state $\ket{3_{\alpha}}$. Assuming that $t_2$ and
$t_1$ are represented by partially overlapping pulses, each having temporal
width $\tau_{\rm w}$, the adiabaticity condition (\ref{adiabat}) requires 
$t_{1,2}^{\rm max} \tau_{\rm w} \gg 1$. 

\begin{figure*}[t]
\centerline{\includegraphics[width=12cm]{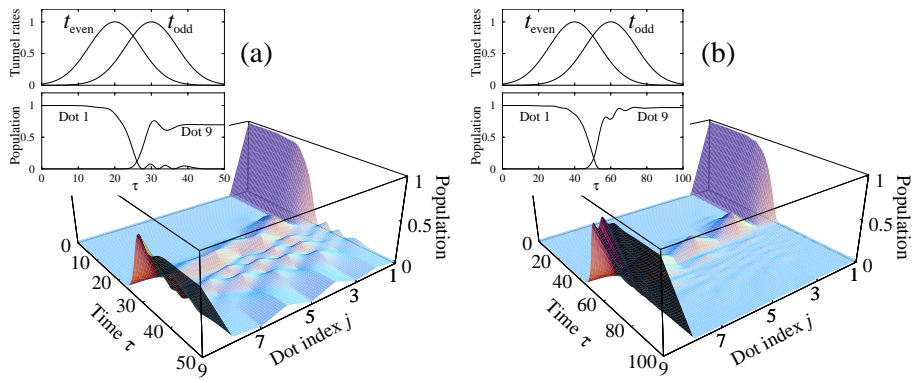}}
\caption{Time-evolution of a single-electron wavepacket in a chain of 
$N=9$ QDs with time-dependent, counterintuitive tunneling rates. 
(a)~Population transfer is incomplete, $|A_N|^2 \simeq 0.7$, when the 
adiabatic condition is not very well satisfied. 
(b)~Almost complete population transfer is achieved, $|A_N|^2 \simeq 0.97$, 
when the adiabatic condition is better satisfied by doubling the temporal
widths of the pulses and the total interaction time (note the different 
scales of the time axis in (a) and (b)). The insets show the 
time-dependence of even and odd tunneling rates and the 
populations of the first and last QDs.
\label{qdc_1e_Dyn}}
\end{figure*} 

In the field of atomic/molecular physics, this technique, involving the 
so-called counterintuitive sequence of pulses, is known as the stimulated
Raman adiabatic passage (STIRAP) that is commonly used for coherent 
population transfer in three-state systems \cite{stirap-rev}. We note
that the solid-state implementations of the CPT and STIRAP in a pair 
of coupled quantum dots driven by two electromagnetic fields has been 
proposed in \cite{QDsOptCPT}. The single electron transfer in a chain
of three QDs via counterintuitive pulsing of tunnel-couplings as 
discussed above has been studied by Greentree et al. in \cite{GCHH}, 
where it was termed coherent tunneling by adiabatic passage (CTAP). 
These authors also considered the extension of CTAP to multidot systems
employing the so-called straddling scheme of \cite{stirapN-DT}.
Other schemes for adiabatic electron transport in tunnel-coupled 
QDs have been discussed in \cite{ETrAdabat}.

Another extension of the STIRAP technique to systems containing more 
than just three states has been given in \cite{stirapNsqLs}. This
scheme can easily be adapted to our system, as described below. We 
thus consider a chain of $N$ sequentially coupled QDs and assume that 
the individual tunnel couplings can selectively and independently be 
manipulated. When $N$ is odd, i.e. $N=3,5,7,\ldots$, the Hamiltonian
(\ref{Ham1ealpha}) has a CPT eigenstate 
\begin{eqnarray}
\ket{\psi_0^{\alpha}} &=& \frac{1}{\sqrt{\mathcal{N}_0}} 
[t_2 t_4 \ldots t_{N-1} \ket{1_{\alpha}} +
(-1) t_1 t_4 \ldots t_{N-1} \ket{3_{\alpha}}  
\nonumber \\ & & \quad + \ldots 
+ (-1)^J t_1 t_3 \ldots t_{N-2} \ket{N_{\alpha}}], \label{CPTNl} \\
& & J \equiv  \frac{1}{2} (N-1) , \nonumber  
\end{eqnarray}
with eigenvalue $\la_0 =0$. Thus the amplitude of the initial state
$\ket{1_{\alpha}}$ is proportional to the product of all the even-numbered
tunnel-couplings, while the amplitude of state $\ket{N_{\alpha}}$ is 
given by the product of all odd-numbered tunnel-couplings, divided by
the normalization parameter $\mathcal{N}_0 = (t_2 t_4 \ldots t_{N-1})^2
+ \ldots + (t_1 t_3 \ldots t_{N-2})^2$. Therefore, if all the even-numbered
tunnel-couplings are pulsed together first, the CPT state (\ref{CPTNl}) 
would coincide with the initial state $\ket{1_{\alpha}}$.
This is then followed by switching-on all the odd-numbered tunnel-couplings,
while the even-numbered ones decrease, which will result in a complete
transfer of electron wavepacket to the state $\ket{N_{\alpha}}$.
If we assume that these two families of pulses are described by common
shape functions, $t_2, t_4, \ldots , t_{N-1} = t_{\rm even}$ and
$t_1, t_3, \ldots , t_{N-2} = t_{\rm odd}$, Eq. (\ref{CPTNl}) takes 
a compact form
\begin{eqnarray}
\ket{\psi_0^{\alpha}} &=& \frac{1}{\sqrt{\mathcal{N}_0}} 
\sum_{n=0}^{J} (-t_{\rm odd})^n \, 
t_{\rm even}^{J-n} \, \ket{(2n+1)_{\alpha}}, \label{CPTNlcmp} \\
& & \mathcal{N}_0 = \sum_{n=0}^{J} t_{\rm odd}^{2n} \, 
t_{\rm even}^{2(J-n)} , \nonumber
\end{eqnarray}
which makes the above discussion more transparent. In particular, complete 
population transfer from the initial state $\ket{1_{\alpha}}$ to the final 
state $\ket{N_{\alpha}}$ can be achieved by applying first the 
$t_{\rm even}$ pulses and then the $t_{\rm odd}$ pulses, the two sets 
of pulses partially overlapping in time, as shown in Fig.~\ref{qdc_1e_Dyn}. 

In order to minimize the nonadiabatic coupling of the CPT state to  
other eigenstates of the system, the rate of change of $t_{\rm even}$ 
and $t_{\rm odd}$, given approximately by the inverse pulse-width 
$\tau_{\rm w}^{-1}$, should be small compared to corresponding 
eigenenergies $|\la| \sim |t_{\rm even} + t_{\rm odd}|$, which yields 
the same condition as above, $t_{\rm even, odd}^{\rm max} \tau_{\rm w} \gg 1$. 
One can see from the results in Fig.~\ref{qdc_1e_Dyn}(a), which were
obtained precisely for this reason, that when this condition is not
very well satisfied, the population transfer is incomplete. As expected,
when the tunneling rates are pulsed for longer times, or, equivalently, 
have larger amplitudes, the adiabaticity condition is satisfied better,
resulting in the complete population transfer from the initial to the 
final dot of the chain, as seen in Fig.~\ref{qdc_1e_Dyn}(b). The 
remarkable advantage of this method over the one described in the 
previous Section is that as long as the two sets of partially overlapping
pulses are strong enough, the adiabatic transfer of population is expected
to be robust with respect to small uncertainties and fluctuations of 
tunneling rates, just like its atomic/molecular counterpart in 
Refs. \cite{stirap3ls,stirap-rev,stirapNsqLs}. On the other hand, the 
electron transfer via effective collective $\pi$ pulse can be achieved 
with smaller tunneling rates and/or reduced interaction times, provided 
a precise control of the tunneling amplitudes and timings is possible. 
Depending on the characteristics of the particular system, one or the 
other method may prove to be more practical.

\section{Conclusions}
\label{sec:concl}

In the above Sections, we have studied the dynamics of a single-electron
transport in a linear array of tunnel coupled quantum dots. We have 
identified two regimes under which a complete coherent transfer of 
electron wavepacket between the two ends of the array can be achieved. 
Our results could be used for reliable information exchange between 
distant parts of an integrated quantum computer. As already noted
in the Introduction, one of the difficulties with the existing proposals 
for integrated QD based QCs \cite{LDV,QDQCdsgn} is that the qubits 
(electron spins) interact with the nearest neighbors only, and there 
is no efficient way of transferring the information between distant 
qubits. As a way around such difficulties, one can envision an integrated
quantum register composed of a large number of sub-registers, each
containing two or more adjacent qubits, represented by spins of single
electrons in individual QDs. The sub-registers are embedded in a
two-dimensional array of empty QDs. As we have shown in an earlier
publication \cite{weNPL}, through the mechanism of transient Heisenberg
coupling, combined with the control of tunnel-coupling between the dots
studied in this paper, this two-dimensional grid could realize a flexible
quantum channel, capable of connecting any pair of qubits within the 
register. Thus, to transfer the information, one connects distant 
sub-registers by a chain of empty QDs and applies one of the protocols
described in the previous Sections to achieve a non-dispersive transfer
of the qubit, followed by its controlled entanglement with a target qubit
\cite{LDV}. Note that this scheme is analogous to a proposal for an 
integrated ion trap based QC \cite{ingrIT}, where, in order to circumvent
the difficulties associated with a single large ion trap quantum register, 
it has been proposed to use many small sub-registers, each containing only 
a few ions, and connect these sub-registers to each other via controlled 
qubit (ion) transfer to the interaction region (entangler) represented 
by yet another ion trap. 

We should note that the coherent electron dynamics in arrays of 
tunnel-coupled QDs bears many analogies with spin-wave dynamics in 
spin chains \cite{EckBose} or electromagnetic field dynamics in periodic
photonic crystals \cite{mher,wgarray}, where some of the effects 
described above should be observable. With an unprecedented control 
over system parameters, arrays of QDs doped with more than one electron
allow for studies of numerous coherence and correlation effects in 
many-body physics.

\begin{acknowledgments}
This work is an outgrowth of earlier collaborative work with
Dr. G.M.~Nikolopoulos which we gratefully acknowledge.
\end{acknowledgments}


\begin{thebibliography}{99}
\bibitem{cmnt} 
Typically, in $\sim 50$ nm size GaAs/AlGaAs QDs, separated from each 
other by $\sim 100$ nm, one has $t_j \sim 0.05$ meV, $\De \veps \sim 0.4$ meV,
$U \sim 10$ meV, and at dilution-refrigerator temperatures $T \sim 2-10$ mK 
the thermal energy is $k_{\rm B} T \sim 0.2-1$ $\mu$eV \cite{QDrev}.
\end{thebibliography}
\end{document}